\documentstyle[prl,floats,multicol,amssymb,amsgen,amsfonts,amsbsy,aps]{revtex}

\tighten

\newcommand{\beq}{\begin{equation}}
\newcommand{\eeq}{\end{equation}}
\newcommand{\beqa}{\begin{eqnarray}}
\newcommand{\eeqa}{\end{eqnarray}}
\newcommand{\ket} [1] {\vert #1 \rangle}
\newcommand{\bra} [1] {\langle #1 \vert}

\newcommand{\mean}[1]{\langle #1 \rangle}
\newcommand{\commut}[2]{[ #1 , #2 ]}

\begin{document}
\title{Phase conjugation of continuous quantum variables}
\author{N. J. Cerf$^{1,2}$ and S. Iblisdir$^1$}
\address{$^1$ Ecole Polytechnique, CP 165, Universit\'e Libre de Bruxelles,
B-1050 Bruxelles, Belgium\\
$^2$ Jet Propulsion Laboratory,
California Institute of Technology, Pasadena, CA 91109\\}
\date{December 2000}
\draft
\maketitle

\begin{abstract}
The phase conjugation of an unknown Gaussian state 
cannot be realized perfectly by any physical process. A semi-classical argument
is used to derive a tight lower bound on the noise that must be introduced 
by an approximate phase conjugation operation. A universal
transformation achieving the optimal imperfect phase conjugation 
is then presented, which is the continuous counterpart 
of the universal-{\sc not} transformation for quantum bits.
As a consequence, it is also shown that more information can be encoded
into a pair of conjugate Gaussian states than using twice the same state.
\end{abstract}


\begin{multicols}{2}

The {\em spin-flip} operation cannot be performed on an arbitrary
spin-1/2 particle (or qubit) since it is an {\em anti\-unitary}
transformation. In other words, given a spin-1/2 particle polarized in
an unknown direction $\vec n$, the state $\ket{\vec n}$ cannot be
turned into $\ket{-\vec n}$ by any physical process. 
Recently, however, it has been shown that
this operation can be done {\em imperfectly}, with a same fidelity for all
states $\ket{\vec n}$, by using a universal quantum spin-flip (or
universal-{\sc not}) transformation\cite{bib_buzek,bib_gisin}. This
transformation yields $\ket{-\vec n}$ with a fidelity of 2/3, which,
remarkably, coincides with the fidelity of the optimal measurement of
a spin-1/2 particle\cite{bib_massar}. This means that the optimal
spin-flip operation can be achieved by first measuring the spin in an
arbitrary direction, then preparing a spin state pointing in the
opposite direction to the measured spin. A related result is that
encoding a space direction $\vec n$ into two antiparallel spins
$\ket{\vec n,-\vec n}$ is slightly more efficient than using a
naive encoding with parallel spins $\ket{\vec n,\vec n}$~\cite{bib_gisin}.

In this paper, we investigate the continuous analogue of the spin-flip
operation, namely the {\em phase conjugation} (or, equivalently,
time reversal). First, we analyze the impossibility
of perfectly conjugating an arbitrary Gaussian state 
(or, in particular, a coherent state $\ket{\alpha}$).
We find that such a process necessarily effects a noise that
is equal to at least {\em twice} the vacuum fluctuation noise
of the input coherent state. 
This leads us to define a universal phase conjugator
or universal-{\sc not} operator for
continuous quantum variables. We then show 
that this transformation is optimal as it achieves 
the lower bound derived above.
The resulting phase conjugation fidelity is 1/2, which, just as
for qubits, is the same as the fidelity of the optimal measurement of
a coherent state\cite{bib_arthurs,bib_stenholm,bib_iblisdir}. Finally, the
link with quantum state estimation and quantum cloning 
for coherent states is discussed. In particular,
it is shown that, in analogy with the situation for qubits,
it is more efficient to encode information
into a pair of conjugate coherent states $\ket{\alpha}\otimes \ket{\alpha^*}$
rather than using twice the same state $\ket{\alpha}^{\otimes 2}$.
The error variance on the real and imaginary parts of $\alpha$
can actually be divided by {\em two} in the former case 
(by applying an appropriate {\em entangled} measurement), 
with respect to the latter case.

Consider a single mode of the electromagnetic field, denoted as $\hat
a=(\hat x+i\hat p)/\sqrt{2}$. The phase conjugation operation consists
in flipping the sign of quadrature $\hat p$ while keeping 
quadrature $\hat x$ unchanged,
that is, replacing $\hat a$ by its Hermitian conjugate $\hat
a^\dagger$. Clearly, this operation is impossible as it does not
conserve the commutation relation: if $\hat b=\hat a^\dagger$ is the
resulting mode, we have $[\hat b,\hat b^\dagger]=-[\hat a,\hat
a^\dagger] =-1$ instead of $1$ ($\hbar=1$). A semi-classical argument
can be used to show that this operation cannot be performed with an
added noise that is lower than a minimum equal to twice the vacuum
noise. Let us consider two modes (mode 0 and 1) that are initially
prepared in the Einstein-Podolsky-Rosen (EPR) state, that is, the
common eigenstate of $\hat X=\hat x_0-\hat x_1$ and $\hat P=\hat
p_0+\hat p_1$ with zero eigenvalue for both operators $\hat X$ and
$\hat P$. Since $[\hat X,\hat P]=0$, these operators can be
diagonalized simultaneously, so that the EPR state can be understood
as representing two particles with a relative position $x_0-x_1$ and a
total momentum $p_0+p_1$ both arbitrarily close to zero. Assume now
that we apply a phase conjugation operator on mode 1, that is, $\hat
x_1'=\hat x_1$ and $\hat p_1'=-\hat p_1$, while mode 0 is left
unchanged. The EPR state is then transformed into the common
eigenstate with zero eigenvalue of operators $\hat X'$ and $\hat P'$,
defined as
\beqa
\hat x_0 - \hat x_1 &=& \hat x_0' - \hat x_1'\equiv \hat X',  \nonumber\\
\hat p_0 + \hat p_1 &=& \hat p_0' - \hat p_1'\equiv \hat P'.   
\eeqa
Importantly, $\hat X'$ and $\hat P'$ cannot commute any more here if
the transformed modes 0' and 1' are to obey the standard commutation relations,
so it is indeed impossible to obtain a common eigenstate of $\hat x_0'
- \hat x_1'$ and $\hat p_0' - \hat p_1'$. Instead, since $\commut{\hat
X'}{\hat P'}=\commut{\hat x_0'}{\hat p_0'}+\commut{\hat x_1'}{\hat
p_1'}=2i$, the Heisenberg uncertainty relation implies that
\beq   \label{eq_heisenberg}
\Delta\hat X' \; \Delta\hat P' \ge {1\over 2} 
|\mean{\commut{\hat X'}{\hat P'}}|= 1.
\eeq
If we now assume that the phase conjugation process introduces some
noise, then it is easy to determine the minimum amount of such noise for
the Heisenberg uncertainty relation to be satisfied. Let us suppose
that mode 1 suffers, after phase conjugation, from a random noise
$n_x$ and $n_p$ on quadrature $\hat x_1'$ and $\hat p_1'$,
respectively. Thus, $\hat x_1'=\hat x_1 + n_x$ and $\hat p_1'=-\hat
p_1 + n_p$. Naturally, we assume that this noise is unbiased, that is,
$\mean{n_x}=\mean{n_p}=0$. Since we are seeking for a ``universal''
transformation, we require the variances of $n_x$ and $n_p$ to be the
same ($\mean{n_x^2}=\mean{n_p^2}=\sigma^2$). The resulting variance of
operators $\hat X'=\hat x_0 - \hat x_1 -n_x$ and $\hat P'=\hat
p_0+\hat p_1-n_p$ is
\beq   \label{eq_variance}
\Delta \hat X'^2 = \Delta \hat P'^2 = \sigma^2,
\eeq
since $\hat x_0 - \hat x_1$ and $\hat p_0+\hat p_1$ have both a
vanishing variance in the EPR state. 
Equation~(\ref{eq_heisenberg}) then implies that
\beq  \label{bound}
\sigma^2 \geq 1,
\eeq
so that the noise induced by the phase conjugation process is lower
bounded by 1, i.e., {\em twice} the variance of a quadrature 
in the vacuum state ($\Delta x^2_{\rm vac}=1/2$).

Let us now construct an actual phase-conjugating transformation that
attains this bound. The input mode, assumed to be prepared in a
coherent state $\ket{\alpha}$, is coupled to an ancilla mode by some
unitary transformation. Subsequently, the ancilla is traced over, so
the processed mode is left in a mixed state that is required to be as
close as possible to the complex conjugate state $\ket{\alpha^*}$. Let
us denote the input mode by $\hat a_1$ and the ancilla mode by $\hat a_2$. 
The canonical transformation can be generally described as
\beq\label{eq:lct}
\hat b_i=M_{ij} \hat a_j + L_{ij} \hat a_j^\dagger,
\eeq
where $i,j=1,2$, and the sum is implicit. 
The output modes $\hat b_1$ and $\hat b_2$
refer to the phase-conjugator output and the processed ancilla, respectively. 
This transformation is determined, in general, by 8 complex
coefficients, but we will now impose the constraints for it
to characterize an (imperfect) phase conjugator.
First, we note that it is always possible to perform a phase
transformation $\hat a_i \to e^{i\phi_i} \hat a_i$ 
and $\hat b_i \to e^{i\psi_i} \hat b_i$
such that the coefficients $M_{1j}$ and $L_{1j}$ are real and
positive. Then, by definition, we require that the phase conjugator obeys
$\mean{\hat b_1}=\mean{\hat a_1^{\dagger}}$. 
Also, without loss of generality, we can
assume that the ancilla is initially in the vacuum state
$\mean{\hat a_2}=\mean{(\hat a_2)^2}=0$ (see \cite{bib_caves}). 
Thus, we must have $M_{11}=0$ and $L_{11}=1$. 
We now impose the ``universality'' of the
transformation, that is, the constraint that the added noise
is phase-insensitive (each quadrature suffers from a the same noise).
If the input mode has phase-insensitive noise, i.e., if
$\mean{(\hat a_1)^2}=\mean{\hat a_1}^2$ 
(for example, if it is a coherent state),
then we require that the output mode also has phase-insensitive noise,
i.e., $\mean{(\hat b_1)^2}=\mean{\hat b_1}^2$. Using
\beq
\mean{(\hat b_1)^2}-\mean{\hat b_1}^2=
\mean{(\hat a_1^\dagger)^2}-\mean{\hat a_1^\dagger}^2
+M_{12}L_{12}
\eeq
we conclude that the universality condition is simply $M_{12}L_{12}=0$.
Three more conditions come from imposing the commutation
rules to be conserved by the transformation (\ref{eq:lct}):
\beqa
\commut{b_1}{b_1^{\dagger}}&=&M_{12}^2-L_{12}^2-1=1, \label{eq1}\\ 
\commut{b_2}{b_2^{\dagger}}&=&|M_{21}|^2+|M_{22}|^2-|L_{21}|^2-|L_{22}|^2=1, 
\label{eq2} \\
\commut{b_1}{b_2}&=&M_{1j}L_{2j}-L_{1j}M_{2j}=0  \label{eq3}. 
\eeqa
Equation~(\ref{eq1}), together with the universality condition, implies
that $L_{12}=0$ and $M_{12}=\sqrt{2}$. Equations (\ref{eq2}) and (\ref{eq3})
then impose two last conditions on the four coefficients $M_{2j}$ and $L_{2j}$,
so we are left with two free parameters.
If we further impose that mode 2 transforms just as mode 1
($M_{22}=0$ and $L_{22}=1$), then we get
\beqa
\hat b_1&=&\hat a_1^{\dagger}+\sqrt{2} \; \hat a_2, \\
\hat b_2&=&\sqrt{2} \; \hat a_1+\hat a_2^{\dagger}.
\eeqa

As we could expect, this transformation exactly describes a
phase-insensitive phase-conjugating linear amplifier (see \cite{bib_caves}).
One can easily check that the noise variance of the output of this
phase conjugator is
\beq
(\Delta x^2)_{b_1}=(\Delta p^2)_{b_1} 
=\Delta x^2_{\rm vac}+2\Delta x^2_{\rm vac}=3/2
\eeq
so that the phase-conjugation induced noise
is {\em twice} the vacuum noise, i.e., $2 \Delta x^2_{\rm vac}=1$.
Hence, this transformation is optimal 
as it saturates the bound~(\ref{bound}). In particular, if the input
is a coherent state $\ket{\alpha}$, the output will be a Gaussian
mixture of coherent state $\rho$ with variance one 
centered on $\ket{\alpha^*}$. Consequently, the phase conjugating fidelity is 
\beq
F=\bra{\alpha^*} \rho \ket{\alpha^*}=1/2,
\eeq
just as for an optimal
measurement\cite{bib_arthurs,bib_stenholm,bib_iblisdir}. 
Interestingly, this implies that phase conjugation is intrinsically a classical
process. It could be achieved as well by simultaneously measuring 
the two quadratures of $\ket{\alpha}$,
and then preparing a coherent state whose quadrature $p$ has a flipped sign.
Incidentally, we note that any number of phase-conjugated outputs
can actually be prepared together at no cost (with $F=1/2$ for each).

It is interesting, at this point, to extend the parallel with
the universal quantum spin-flip machine for qubits, and make a connection
with a state estimation question. In \cite{bib_gisin},
Gisin and Popescu have found the surprising result 
that encoding a direction $\vec{n}$ into
two antiparallel spins $\ket{\vec n, -\vec n}$ 
yields slightly more information on $\vec{n}$ than
encoding it into two parallel spins $\ket{\vec n,\vec n}$.
Here, we investigate the counterpart of this situation
for information that is carried by a continuous quantum variable
instead of a qubit. Consider the situation where Alice 
wants to communicate to Bob a complex number $\alpha=(x+ip)/\sqrt{2}$. 
Assume Alice is allowed to use a quantum channel
only twice so as to send Bob two coherent states of a given
amplitude $|\alpha|^2$ each.
She can choose, for example, 
to send Bob the product state $\ket{\alpha}^{\otimes 2}$.
In this case, the best strategy to infer both $x$ and $p$ 
with a same precision is to perform 
a product measurement\cite{bib_stenholm}.
A simultaneous measurement of the two
quadratures of each coherent state $\ket{\alpha}$ yields $(x,p)$
with a variance $2 \Delta x^2_{\rm vac}=1$~\cite{bib_arthurs}.
The resulting error variance on $x$ and $p$ estimated from these two
measurements is then equal to one half of this variance,
that is $\Delta x^2_{\rm vac}=1/2$. (This is just the statistical factor.)

Another possibility is that Alice sends Bob the product 
state $\ket{\alpha}\otimes \ket{\alpha^*}$.
In this case, a possible (but not necessarily optimal)
strategy for Bob is again to carry out a product measurement, 
taking into account that
the measured value of $p$ of the second state 
should be read as $-p$. This obviously
results in the same error variance 1/2. However, the fact that the
continuous universal-{\sc not} transformation has a 
non-unity fidelity leaves open the possibility that there exists
a measurement of $\ket{\alpha}\otimes \ket{\alpha^*}$
that is {\em not} of a product form, and yields a variance strictly lower
than 1/2. Indeed, if there was a perfect universal phase conjugator,
then it could be used to convert $\ket{\alpha^*}$ into $\ket{\alpha}$
before applying the optimal product measurement on $\ket{\alpha}^{\otimes 2}$,
thereby resulting in the same minimum variance in both cases.

Let us now explicitly describe an {\em entangled} measurement
of the product state $\ket{\alpha}\otimes \ket{\alpha^*}$, 
which yields indeed a lower variance. Expressing the two input modes
as $\ket{\alpha}=\exp(ip \hat x_1 -ix \hat p_1)\ket{0}$
and $\ket{\alpha^*}=\exp(-ip \hat x_2 -ix \hat p_2)\ket{0}$,
we can write the input product state as
$\ket{\alpha}\otimes \ket{\alpha^*}= \exp(ip \hat X -ix \hat P) \ket{0}$,
where $\hat X=\hat x_1 - \hat x_2$ and $\hat P=\hat p_1 + \hat p_2$
are two {\em commuting} operators. Assume now that 
the two input states $\ket{\alpha}$ and $\ket{\alpha^*}$ are sent
each into one of the inputs of a balanced beam splitter, characterized
by the canonical transformation
\beqa
\hat x_1' &=& (\hat x_1+\hat x_2)/\sqrt{2},  
\qquad  \hat p_1' = (\hat p_1+\hat p_2)/\sqrt{2}, \\
\hat x_2' &=& (\hat x_1-\hat x_2)/\sqrt{2},  
\qquad  \hat p_2' = (\hat p_1-\hat p_2)/\sqrt{2}. 
\eeqa
The input product state can be reexpressed as
\beq
\ket{\alpha}\otimes \ket{\alpha^*}= \exp(i\sqrt{2}\;p\; \hat x_2' 
-i\sqrt{2}\; x\; \hat p_1') \ket{0}
\eeq
implying that $x$ and $p$ can be measured {\em separately} here by applying
homodyne detection on modes 1' and 2'. Indeed,
a measurement of the first quadrature of mode 1'
yields $\sqrt{2}\, x$, on average, while a measurement 
of the second quadrature of mode 2' yields $\sqrt{2}\, p$. 
These two measurements suffer each from an error
of variance $\Delta x^2_{\rm vac}=1/2$. Hence, the resulting error variance
on $x$ and $p$ is reduced to $\Delta x^2_{\rm vac}/2=1/4$.
In contrast, if we had the input product state $\ket{\alpha}^{\otimes 2}$ and
were sending each coherent state $\ket{\alpha}$ into an input of
a balanced beam splitter, we would obtain a single coherent state 
$\ket{\sqrt{2}\; \alpha}$ on output mode 1'. One should then necessarily
perform a {\em simultaneous} measurement of the two quadratures 
of the latter mode, yielding $(\sqrt{2}\;x,\sqrt{2}\;p)$ with
an error variance $2 \Delta x^2_{\rm vac}=1$, or, equivalently
$x$ and $p$ with a variance $\Delta x^2_{\rm vac}=1/2$.
As a consequence, we have proven here that a better strategy 
for sending $x$ and $p$ to Bob is to encode them
into two conjugate coherent states $\ket{(x+ip)/\sqrt{2}}\otimes
\ket{(x-ip)/\sqrt{2}}$ 
rather than sending two replicas of $\ket{(x+ip)/\sqrt{2}}$. The error
variance on $x$ and $p$ is indeed reduced by a factor of two
via the use of phase conjugation.

Finally, let us discuss the connection between the universal phase conjugator
and quantum cloning. It can be shown that the Gaussian cloning machine for
continuous variables introduced in \cite{bib_cerf} generates, in
addition to the two clones of the input state, an imperfect
phase-conjugate version of the input state with the same fidelity
($F=1/2$) as that of the universal phase-conjugator\cite{bib_braunstein}. 
The exact same
property holds for the universal qubit cloner \cite{bib_buz_hil},
which also yields a flipped qubit with a fidelity equal to that of the
universal quantum spin-flip machine\cite{bib_gisin}. Thus, the general
rule seems to apply that the production of two clones is necessarily
accompanied by the creation of one anticlone (time-reversed state).

As a last comment, it is worthwhile noting that we have here another example
of the classical nature of the universal-{\sc not} operation. As emphasized
in \cite{bib_gisin}, spin flipping is essentially a classical
operation on qubits, since it can be done by a measurement followed
by the preparation of a flipped spin. This also implies that
any number of flipped spins can be produced together with the same fidelity.
Similarly, we have shown here that the same situation prevails
for the phase conjugation of continuous quantum variables. It seems
therefore tempting to conjecture that any (imperfect) time-reversal procedure
can be done optimally in a classical way. Proving this conjecture
and understanding the fundamental reason for it are interesting
open questions.

We are grateful to Serge Massar for useful discussions.
S. I. acknowledges support from the Fondation Universitaire Van Buuren
at the Universit\'e Libre de Bruxelles.

\end{multicols}

\begin{references}

\vspace{-1cm}

\bibitem{bib_buzek} V. Buzek, M. Hillery, and R. F. Werner,
 Phys. Rev. A {\bf 60}, R2626 (1999).
\bibitem{bib_gisin} N. Gisin and S. Popescu, Phys. Rev. Lett. {\bf 83},
 432 (1999).
\bibitem{bib_massar} S. Massar and S. Popescu, Phys. Rev. Lett. {\bf 74},
 1259 (1995).
\bibitem{bib_arthurs} E. Arthurs and J. L. Kelly, Jr., Bell Syst. Tech. J.
 {\bf 44}, 725 (1965). 
\bibitem{bib_stenholm} S. Stenholm, Ann. Phys. (N.Y.) {\bf 218}, 233 (1992).  
\bibitem{bib_iblisdir} N. J. Cerf and S. Iblisdir, Phys. Rev. A {\bf 62},
 040301 (2000).
\bibitem{bib_caves} C. M. Caves, Phys. Rev. D {\bf 26}, 1817 (1982).
\bibitem{bib_cerf} N. J. Cerf, A. Ipe, and X. Rottenberg, Phys. Rev. Lett.
{\bf 85}, 1754 (2000).
\bibitem{bib_braunstein} S. L. Braunstein, N. J. Cerf, S. Iblisdir,
 P. van Loock, and S. Massar, in preparation.
\bibitem{bib_buz_hil} V. Buzek and M. Hillery, Phys. Rev. A {\bf 54}, 
 1844 (1996).

\end{references}
\end{document}